\documentclass[twocolumn,showpacs,preprintnumbers,nofootinbib]{revtex4} 
\usepackage{graphicx}
\usepackage{latexsym}
\usepackage{amssymb}
\def\lsim{\mathrel{\rlap{\lower4pt\hbox{\hskip1pt$\sim$}}
    \raise1pt\hbox{$<$}}}
\def\gsim{\mathrel{\rlap{\lower4pt\hbox{\hskip1pt$\sim$}}
    \raise1pt\hbox{$>$}}}
\def\sqr#1#2{{\vcenter{\vbox{\hrule height.#2pt
         \hbox{\vrule width.#2pt height#1pt \kern#1pt
         \vrule width.#2pt}
         \hrule height.#2pt}}}}

\def\beq{\begin{equation}}
\def\eeq{\end{equation}}
\def\beqa{\begin{eqnarray}} 
\def\eeqa{\end{eqnarray}}
\begin{document}

\title{Time Evolution of the Fine Structure Constant in a Two-Field
Quintessence Model}

\author{M. C. Bento}

\altaffiliation[Also at ] {CFIF, Instituto Superior T\'ecnico, Lisboa. 
Email address: bento@sirius.ist.utl.pt}

\author{O. Bertolami}

\altaffiliation[Also at ] {CFN, Universidade de Lisboa. Email address:
orfeu@cosmos.ist.utl.pt}

\author{N. M .C. Santos} 

\altaffiliation[Also at ] {CFIF, Instituto Superior T\'ecnico, Lisboa.
 Email address: ncsantos@cfif.ist.utl.pt }

\affiliation{ Departamento de F\'\i sica, Instituto Superior T\'ecnico \\
Av. Rovisco Pais 1, 1049-001 Lisboa, Portugal}

\vskip 0.5cm

\date{\today}

\begin{abstract}
 
We examine the variation of the fine structure constant in the context
of a two-field quintessence model. We find that, for solutions
that lead to a transient late period of accelerated expansion, it is
possible to fit the data arising from quasar spectra and comply with
the bounds on the variation of $\alpha$ from the Oklo reactor,
meteorite analysis, atomic clock measurements, Cosmic Microwave 
Background Radiation and Big Bang Nucleosynthesis. That is more 
difficult if we consider solutions corresponding to a late 
period of permanent accelerated expansion.

\vskip 0.5cm
 
\end{abstract}

 \pacs{98.80.-k,98.80.Cq,12.60.-i\hspace{4cm} Preprint DF/IST-1.2004}

\maketitle
\vskip 2pc
\section{Introduction}

The recent claim that the spectra of quasars (QSOs) indicates the
variation of the fine structure constant, $\alpha$, on cosmologically
recent times
\cite{Murphy:2003hw,Murphy:2002ve,Murphy:2001nu,Murphy:2000ns} has
raised considerable interest in examining putative sources of this
variation. In most models a possible variation of the fine structure
constant is studied by arbitrarily coupling fields to
electromagnetism, as suggested by Bekenstein \cite{Bekenstein}.  Thus
the proposals put forward sofar consider a scalar field
\cite{Olive:2001vz,Gardner}  (with an additional coupling to dark
matter and to a cosmological constant in the former case)  and
quintessence \cite{Anchordoqui,Mota}. In fact, as discussed in
Ref. \cite{Kostelecky}, the couplings of gravito-scalar fields, such as the
axion or the dilaton, to electromagnetism naturally arise in $N=4$
Supergravity in four dimensions, making this model particularly
interesting.  It is worth mentioning that, in the latter model, the
mass of these scalars can drive the accelerated expansion of the
Universe, transient or eternal \cite{Bertolami}.

In this work, we shall consider the implications for the variation of
$\alpha$ of a two-field quintessence model \cite{Bento:2001yv}, with
the quintessence fields coupled to electromagnetism, as proposed by
Bekenstein \cite{Bekenstein}. There are several motivations for
studying potentials with coupled scalar fields. Firstly, if one
envisages to describe the Universe dynamics from fundamental theories,
it is most likely that an ensemble of scalar fields (moduli, axions,
chiral superfields, etc) will emerge, for instance, from the
compactification process in string or braneworld
scenarios. Furthermore, coupled scalar fields are invoked for various
desirable features they exhibit, as in the so-called hybrid
inflationary and reheating models \cite{LindeBKB}.  The model
of. Ref.~\cite{Bento:2001yv} has the additional bonus of leading to
transient as well as permanent solutions for the late time
acceleration of the Universe. The former solutions are desirable given
that it has been recently pointed out that an eternally accelerating
universe poses a challenge for string theory, at least in its present
formulation, since asymptotic states are inconsistent with spacetimes
that exhibit event horizons \cite{Hellerman}.  Moreover, it is argued
that theories with a stable supersymmetric vacuum cannot relax into a
zero-energy ground state if the accelerating dynamics is guided by a
single scalar field \cite{Hellerman}, a problem that can be
circumvented in the two-field model we are considering
\cite{Bento:2001yv}.

We now turn to the available observational bounds on the variation of
$\alpha$.  Observations of the spectra of 128 QSOs with $z=0.2 - 3.7$
suggest that, for $z>1$, $\alpha$ was smaller than at present
\cite{Murphy:2003hw,Murphy:2002ve,Murphy:2001nu,Murphy:2000ns}

\beq
{\Delta \alpha \over \alpha} \equiv
{\alpha (z)-\alpha_0 \over \alpha_0} = (-0.54 \pm 0.12) \times
10^{-5}~,
\label{Murphy}
\eeq
at $4.7 \sigma$. 

The most recent data is from Chand {\it et al.} \cite{Chand1,Chand2}
 obtained via a new sample of Mg~II systems from distant quasars with
 redshifts in the range $0.4 \le z \le 2.3$ yield more stringent
 bounds ($3~\sigma$), namely:

\beq
{\Delta \alpha \over \alpha} = (-0.06 \pm 0.06) \times
10^{-5}~,
\label{Chand1}
\eeq
where terrestrial isotopic abundances have been assumed. If, instead, 
low-metalicity isotopic abundances are assumed, Chand {\it et al.} obtain

\beq
{\Delta \alpha \over \alpha} = (-0.36 \pm 0.06) \times
10^{-5}~,
\label{Chand2}
\eeq in which case the statistical inconsistency with Murphy {\it et
al.}, Eq.~(\ref{Murphy}), is clearly smaller. Notice that, in contrast
with Webb {\it et al.}, who use different lines from different
multiplets and elements, Chand {\it et al.} use mostly Mg II data
yielding a smaller but better quality dataset.

On the other hand, the Oklo natural reactor provides a bound, at $95
\%$ CL,

\beq
-0.9 \times 10^{-7} < {\Delta \alpha \over \alpha} < 1.2
\times 10^{-7}~,
\eeq
 for $z=0.14$
\cite{Damour:1996zw,Fujii:2003gu,Fujii:1998kn}. Notice, however, that 
the use of an
equilibrium neutron spectrum  has been criticized in
Ref.~\cite{Lamoreaux:2003ii}, where a lower bound on the variation of 
$\alpha$ over the last two billion years is given by

\beq
{\Delta \alpha\over \alpha} \geq +4.5 \times 10^{-8}~.
\eeq

Estimates of the age of iron meteorites ($z=0.45$), combined with a
measurement of the Os/Re ratio resulting from the radioactive decay
$^{187}$Re~$\to~^{187}$Os, gives \cite{Olive:2003sq,Fujii:2003uw,
Olive:2002tz} 

\beqa
{\Delta \alpha \over \alpha} = (-8 \pm 8) \times
10^{-7}~, \eeqa at $1\sigma$, and \beqa -24 \times 10^{-7} < {\Delta
\alpha \over \alpha} < 8 \times 10^{-7}~,
\eeqa 
at $2\sigma$.

Notice that, if the variation of the fine structure constant is
linear, the Murphy {\it et al.} observations of QSO absorption spectra and of
geochemical tests from meteorites are incompatible given that the
former yields $\Delta \alpha / \alpha \simeq 5 \times 10^{-6}$ at $z
=0.5$, while the latter leads to $\Delta \alpha / \alpha \le 3 \times
10^{-7}$ at $z=0.45$ \cite{Mota}. However, this problem may not exist
if one uses the Chand {\it et al.} dataset.

Moreover, observations of the hyperfine frequencies of the $^{133}$Cs and
$^{87}$Rb atoms in their electronic ground state, using several laser
cooled atomic fountain clocks, give, at present ($z=0$)
\cite{Marion:2002iw} (see also \cite{Bize:bj})

\beq {\left\vert {\dot{\alpha} \over \alpha}\right\vert} < 4.2 \times
10^{-15}~\mbox{yr}^{-1}~, \eeq 
where the dot represents
differentiation with respect to cosmic time. Tigher bounds arise from
the remeasurement of the $1s-2s$ transition of the atomic hydrogen and
comparison with a previous measurement with respect to the ground
state hyperfine splitting in $^{133}$Cs and combination with the drift
of an optical transition frequency in $^{199}$Hg$^+$, that is
\cite{Fischer}:

\beq
{\dot{\alpha} \over \alpha} =
(-0.9 \pm 4.2) \times 10^{-15}~\mbox{yr}^{-1}~.
\eeq

There are also constraints coming from  Cosmic Microwave
Background Radiation (CMBR), where

\beq
\vert {\Delta \alpha / \alpha} \vert \leq 10^{-2}~,
\eeq
at $z=10^3$ \cite{CMBBBNconstr,CMBconstr}, and from Big Bang
nucleosynthesis
 (BBN), 

\beq
- 6 \times 10^{-4} < {\Delta \alpha / \alpha} < 1.5 \times 10^{-4}~, 
\eeq
at $z=10^8~-~10^{10}$ \cite{CMBBBNconstr,BBNconstr}.

In addition, we should take into account the equivalence principle
experiments, which imply \cite{Olive:2001vz}

\beq 
{\zeta _F \over \sqrt \omega} \leq 10^{-3}~,
\label{zetaf}
\eeq 
where $\zeta _F$ is the
coupling between the scalar and electromagnetic fields and $\omega
\equiv {M_*^2 \over 2 M^2}$,  $M$ being  the reduced Planck
mass $(M\equiv M_P/\sqrt{8\pi})$ and $M_*$ the corresponding analogue 
in the scalar sector.

Notice that not all models put forward sofar satisfy the
abovementioned bounds.  For instance, quintessence models like the
last one in \cite{Anchordoqui} and the $N=4$ Supergravity models in
four dimensions are not consistent with Webb {\it et al.}
data.  In the following, we shall present our
model and show that it is consistent with all the available data for a
suitable choice of the model parameters.

\section{Minimal Coupling of Quintessence to the Electromagnetic Field}

We shall study the time evolution of $\alpha$ in a model where
two homogeneous scalar fields are minimally coupled to
electromagnetism. This evolution follows from the effective action, in
natural units ($M=1$),

\beq
S = \int d^4x \sqrt{-g}~ \left[-\frac{1}{2} R+{\cal L}_b + {\cal L}_Q +
 {\cal L}_{em}\right]~,
\label{action}
\eeq 
where ${\cal L}_b$ is the
Lagrangian density for the background matter (CDM, baryons and
radiation), which we consider homogeneous, corresponding to a
perfect fluid with barotropic equation of state $p_b= w_b~\rho_b$,
with $w_b$ constant ($-1 \leq w_b \leq 1$;
$w_b=1/3$ for radiation and  $w_b = 0$ for dust); 
${\cal L}_Q$ is the Lagrangian
density for the scalar fields

\beq
{\cal L}_{Q} = {1\over2} \partial^{\mu}\phi\partial_{\mu}\phi + 
{1\over2} \partial^{\mu}\psi\partial_{\mu}\psi
- V(\phi,\psi) \,.
\eeq

We consider the two-field quintessence model with potential 
\cite{Bento:2001yv} 

\beq
V(\phi, \psi)=  e^{-\lambda\phi} P(\phi, \psi)~,
\label{pot}
\eeq
where
\beqa
P(\phi, \psi)&=     &      A~+
(\phi-\phi_*)^2+B~(\psi-\psi_*)^2  \nonumber \\
&& ~+ C~\phi(\psi-\psi_*)^2 + D~\psi (\phi-\phi_*)^2~.
\label{pot1}
\eeqa
Potentials of the type exponential times a  polynomial, involving 
one single scalar field, have been considered before, see e.g. \cite{Albrecht}.
The evolution  equations for  a  spatially-flat 
Friedmann-Robertson-Walker  (FRW)
Universe, with Hubble parameter $H\equiv \dot a /a$, are 
\beqa
\dot{H} & = &
- \,   {1\over2}   \left(   \rho_b   +  p_b   +   \dot\phi^2
+\dot\psi^2\right)~, \nonumber      \\
\dot\rho_\gamma      &=& -3H(\rho_b+p_b)~, \nonumber   \\
\ddot\phi    &=& -3H\dot\phi-{\partial_\phi  V}~,\nonumber \\
\ddot\psi  &=&  -  3H\dot\psi  -{\partial_\psi V}~,
\label{fried}
\eeqa 
subject to the Friedmann constraint
\beq
H^2  =  {1\over3}   \left(  \rho_b   +  {1\over2}\dot\phi^2
+{1\over2} \dot\psi^2 +V \right) \ ,
\label{Friedmann}
\eeq
where $\partial_{\phi(\psi)} V  \equiv {\partial V \over \partial
\phi(\psi)}$. 
The total energy density of the homogeneous scalar fields is given
by $\rho_Q=\dot\phi^2/2+\dot\psi^2/2+ V(\phi, \psi)$.
\begin{figure}[t]
\includegraphics[width=8cm]{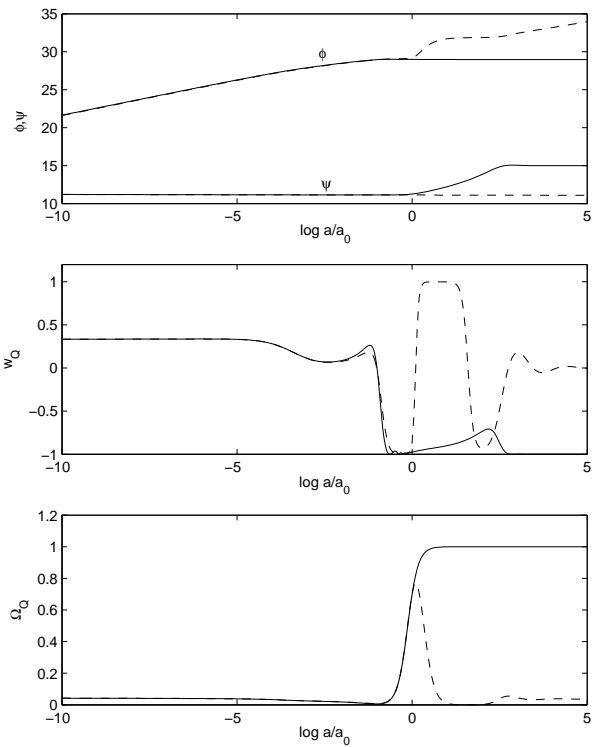}
\caption[fig:fields]{Evolution of the quintessence fields (upper panel), the 
equation of state parameter (middle panel) and the quintessence 
fractional energy density (lower panel), for 
transient (dashed) and permanent (full) acceleration 
solutions.}
\label{fig:fields}
\end{figure}
\begin{figure}[t]
\includegraphics[width=8cm]{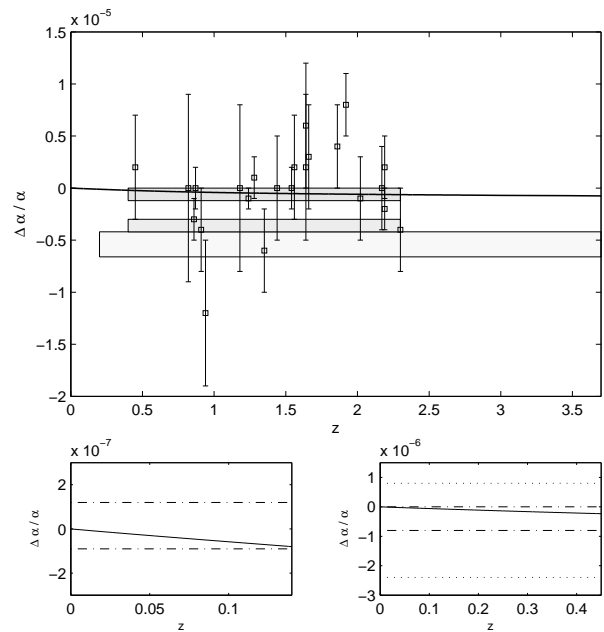}
\caption[fig:transient1]{ Evolution of $\alpha$ for a transient
 acceleration model with $\zeta_1 = 2 \times 10^{-6}$ and $\zeta_2 = 8
 \times 10^{-5}$. In the upper panel, the boxes represent the QSO
 bounds given in Eqs.~(\ref{Chand1}) (top box), (\ref{Chand2}) (middle
 box) and (\ref{Murphy}) (lower box).  Also shown is the QSO
 absorption systems dataset of Chand {\it et al.} (Table 3 of
 Ref. \protect\cite{Chand1}).  The lower panel details the behaviour
 of $\alpha$ for small values of $z$. The lower left plot shows the
 Oklo bound (dash-dotted lines) and the right one the meteorite bounds
 (dash-dotted and dotted lines correspond to $1\sigma$ and $2\sigma$,
 respectively).}
\label{fig:transient1}
\end{figure}
\begin{figure}[t]
\includegraphics[width=8cm]{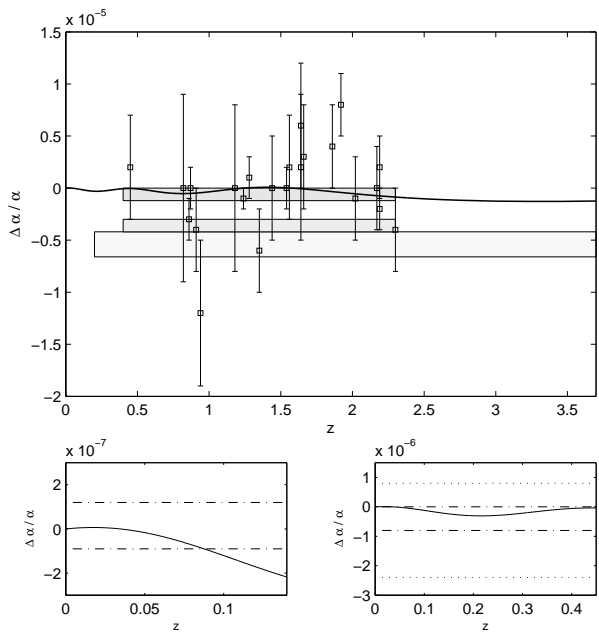}
\caption[fig:permanent1]{As for Fig.~\ref{fig:transient1} but for a permanent
acceleration model with $\zeta_1 = -4 \times 10^{-5}$ and $\zeta_2 = 1
\times 10^{-6}$.}
\label{fig:permanent1}
\end{figure}

Integrating Eqs.~(\ref{fried}), one finds  that
there are essentially two realistic types of solutions, corresponding
to either transient or permanent acceleration, as illustrated in
Figure \ref{fig:fields}.  It is possible to obtain permanent or
transient vacuum domination, satisfying present bounds on observable
cosmological parameters, for a rather broad range of parameters of the
potential, as discussed in \cite{Bento:2001yv}.

Permanent vacuum domination takes place when at least the $\phi$ field
settles at the minimum of the potential; soon after, $\psi$
monotonically increases until it reaches its asymptotic value. For
transient vacuum domination (which occurs either when the potential
has no local minimum or when $\phi$ arrives at the minimum with enough
kinetic energy to roll over the potential barrier) the $\phi$ field
increases monotonically and $\psi$ slightly decreases. The
contribution of $P(\phi,\psi)$, which controls the presence and
location of the minimum,  is only important at recent times,
since at early epochs the exponential factor dominates and the model
behavior depends essentially on parameter $\lambda$. Notice that, as
$\lambda$ increases, the model becomes more similar to $\Lambda$CDM at
early times.

The interaction term between the scalar fields and the electromagnetic
 field is, as suggested in Ref.~\cite{Bekenstein},
   
\beq
{\cal L}_{em} = - \frac{1}{4} B_F(\phi,\psi) F_{\mu\nu}F^{\mu\nu}~,
\label{lem}
\eeq
with a linearly expanded $B(\phi,\psi)$ given by
\beq
B_F(\phi,\psi) = 1-\zeta_1 (\phi-\phi_0)-\zeta_2 (\psi-\psi_0),
\label{BF}
\eeq
where $\phi_0$ and $\psi_0$ are the present values of the scalar fields.
Therefore, the variation of the fine structure constant, $\alpha=
 \alpha_0/B_F(\phi,\psi)$, is given by

\beq
{\Delta \alpha \over \alpha}  =
\zeta_1 (\phi - \phi_0) + \zeta_2  (\psi - \psi_0)\,.
\label{delalpha}
\eeq

Notice that the upper bound on $\zeta_F$, see Eq.~(\ref{zetaf}), 
implies that $\zeta _{1,2} \leq 7 \times 10^{-4}$ because of 
the different definition for $B_F$ used in
\cite{Olive:2001vz}, namely,  $B_F(\varphi) = 1-{\zeta_F \over M_*}
(\varphi-\varphi_0)$.

The  rate of variation of $\alpha$, at present, can be written as

\beq
{\dot{\alpha} \over \alpha} = - \left( \zeta_1 {d \phi\over dy}~+ 
\zeta_2 {d \psi \over dy}\right) H_0 \,,
\label{dotalpha}
\eeq 
where $y \equiv 1 + z$ and $H_0$ is the value of the Hubble
constant at present, $H_0 = (h/9.78)~ \times 10^{-9}~{\rm yr}^{-1}$.

\section{Results}

We adopt the method of choosing the model parameters so as to satisfy
a set of priors for the cosmological parameters ($h = 0.70$,
$\Omega_{m} = 0.3$, $\Omega_{Q} = 0.70$ and $\Omega_{r} = 4.15 \times
10^{-5} h^{-2}$) and then adjust the coupling parameters, $\zeta_1$
and $\zeta_2$, in order to satisfy the bounds on the evolution of
$\alpha$.   Notice that the priors chosen above are
consistent with a combination of WMAP data and other CMB experiments
(ACBAR and CBI), 2dFGRS measurements and Lyman $\alpha$ forest data,
which gives \cite{Spergel:2003cb}: $h=0.71^{+0.04}_{-0.03}$, $\Omega_m
=0.27 \pm 0.04$, $\Omega_{Q} = 0.73 \pm 0.04$ and $w_{Q}<-0.78$
($95\%$ CL), corresponding to the best fit for the observed Universe.

For large $z$, the tightest bound on dark energy arises from
nucleosynthesis, $\Omega_Q(z=10^{10}) < 0.045$, implying that
$\lambda>9$ \cite{Bean}. Considering a possible underestimation of
systematic errors, one gets a more conservative bound, $\Omega_Q<0.09$ at
$2 \sigma$, which corresponds to $\lambda>6.5$.  For the model
parameters chosen in this work (see below), we get $\Omega_Q=0.042$.

The model is also consistent with CMBR observations. Recent CMBR 
data imply that $\Omega_{Q} < 0.39$ at last scattering \cite{Bean}, which
is less stringent than the nucleosynthesis bound and is clearly obeyed
by our model.
We have  also computed the angle subtended by the sound horizon at last
scattering, obtaining
$\theta_A\simeq 0.599^o$, 
clearly within  WMAP bounds on this quantity  \cite{Spergel:2003cb}:
$\theta_A=0.598\pm 0.002^o$.

\begin{figure}[t]
\includegraphics[width=8cm]{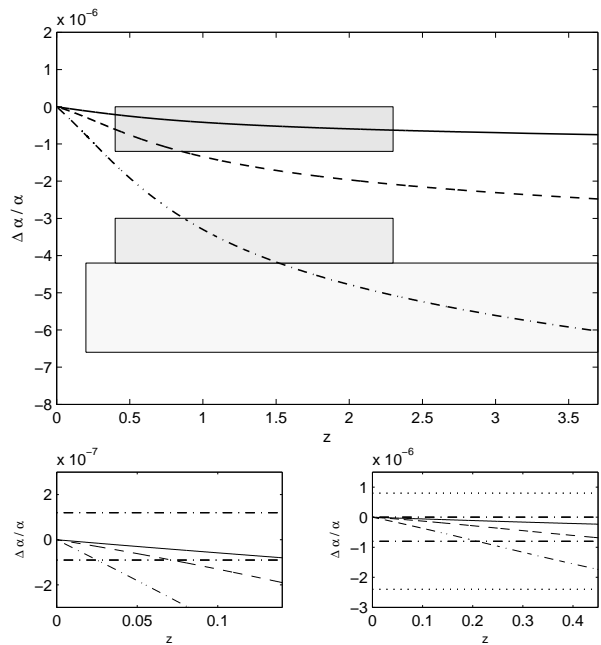}
\caption[fig:transient2]{Evolution of $\alpha$ for a transient
 acceleration model with
$\zeta_1 = 2 \times 10^{-6}$ and $\zeta_2 = 8 \times 10^{-5}$ (full
line), $\zeta_1 = 5.3 \times 10^{-6}$ and $\zeta_2 = 3 \times 10^{-5}$
(dashed line), $\zeta_1 = 1.4 \times 10^{-5}$ and $\zeta_2 = 7 \times
10^{-4}$ (dash-dotted line). Line and box conventions are those of
Figs. \ref{fig:transient1} and \ref{fig:permanent1}.}
\label{fig:transient2}
\end{figure}

\begin{figure}[t]
\centering
 \includegraphics[width=8cm]{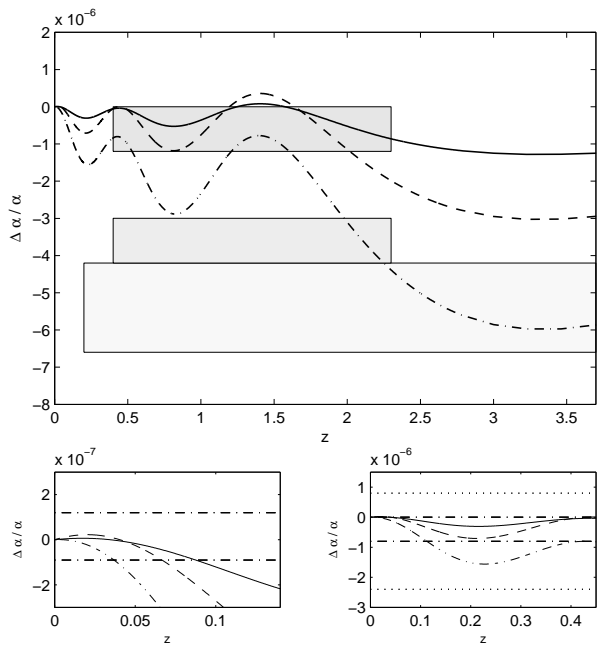}
\caption[fig:permanent2]{As for Fig.~\ref{fig:transient2} but for a permanent
acceleration model with $\zeta_1 = - 4 \times 10^{-5}$ and $\zeta_2 =
1 \times 10^{-6}$ (full line), $\zeta_1 = - 1 \times 10^{-4}$ and
$\zeta_2 = 1 \times 10^{-6}$ (dashed line), $\zeta_1 = - 1.5 \times
10^{-4}$ and $\zeta_2 = 1.4 \times 10^{-5}$ (dash-dotted line).}
\label{fig:permanent2}
\end{figure}

As an example of transient acceleration models, we consider the
following set of parameters: $\lambda=9.5$, $A=0.1$, $B=10^{-3}$, $C =
8 \times 10^{-5}$, $D=2.8$, $\phi_*=28.965$, $\psi_*=20$, with
$\zeta_1 = 2 \times 10^{-6}$ and $\zeta_2 = 8 \times 10^{-5}$.  In
this case, the variation of $\alpha$ is in agreement with both the
Oklo and meteorite bounds, as can be seen in the lower panel of Figure
\ref{fig:transient1}. In the upper panel, we see that the fit to Chand
{\it et al.} QSO data (where terrestrial isotopic abundances have
been assumed) is acceptable: we get $\chi^2/d.o.f = 22/15$ (notice
that, in the computation of the $d.o.f.$, only 8 free variables should
be taken into account since one potential parameter is tuned in order
to have the observed dark energy density at the present). Also the
constraint from atomic clock measurements is verified as we obtain

\beq
{ {\dot{\alpha} \over
\alpha}} = -4.5 \times 10^{-17}~\mbox{yr}^{-1}~.
\eeq   
Moreover, the model
gives variations in $\alpha$ during  BBN and at the last scattering
surface well within the bounds stated earlier, e.g.  we get 

\beqa
&&{\Delta \alpha
\over \alpha}(CMBR) = -2.7 \times 10^{-6}~,\nonumber\\
&&{\Delta \alpha \over
\alpha}(BBN) = -1.1 \times 10^{-6}~.
\eeqa

For models with permanent acceleration, however, consistency with
QSO spectra and Oklo data is  problematic.  Our results are
illustrated in Figure \ref{fig:permanent1}, where we have chosen 
$\lambda=9.5$, $A=0.02$, $B=2 \times 10^{-3}$, 
$C = 6 \times 10^{-4}$, $D=4.5$, $\phi_*=28.9675$,
$\psi_*=15$, for $\zeta_1 = -4 \times 10^{-5}$ and $\zeta_2 =
1 \times 10^{-6}$. We get
\beqa
&&{ {\dot{\alpha} \over
    \alpha}}= 5.2 \times 10^{-17}~\mbox{yr}^{-1}\nonumber\\
&&{\Delta \alpha \over
    \alpha}(CMBR) = 4.5 \times 10^{-5}\nonumber\\
&&{\Delta \alpha \over
    \alpha}(BBN) = 2.9 \times 10^{-4}
\eeqa  
and $\chi^2/d.o.f = 21/15$. Notice that the BBN bound is not respected
in 
this case.

It is clear that the evolution of $\alpha$, past and future, is a
reflection of the dynamics of the fields $\phi$ and $\psi$, which
depends crucially on whether acceleration is permanent or transient.
  In the example
of Figure \ref{fig:transient1}, $\alpha$ is increasing monotonically
and will continue to do so. However if $\zeta_1 \ll \zeta_2$, it is
the $\psi$ field that determines the evolution; hence, in that case,
$\alpha$ may  decrease in the future.  For permanent acceleration
solutions, the sign of $\zeta_2$ determines the future evolution of
$\alpha$, since $\psi$ has more dynamics than $\phi$. Hence, for the
example of Figure \ref{fig:permanent1}, $\alpha$ will increase in the
future since $\zeta_2>0$; the oscillating behavior of $\alpha$
reflects the oscillations of $\phi$ before  it settles at the minimum of
the potential.

In Figs. \ref{fig:transient2} and \ref{fig:permanent2} we show that,
increasing the couplings to the electromagnetic field, $\zeta_1$ and
$\zeta_2$, it is also possible to fit Murphy {\it et al.} QSO data
\cite{Murphy:2003hw}, both in the transient and permanent acceleration
regimes.  However, as  should be expected, it is more
difficult to respect the Oklo and meteorites bounds in both cases.

\section{Conclusions}

In this work, we have studied the variation of the fine structure
constant in the context of a quintessence model with two coupled
scalar fields. We find that transient acceleration models can fit
the latest QSO data and comply with the upper bounds on $\Delta \alpha$
from the Oklo reactor, meteorite analysis and the atomic clock
measurements.  For permanent acceleration models, however, it is more 
difficult to fit the QSO data and satisfy the Oklo, meteorite and BBN
bounds simultaneously.  We have studied the sensitivity of our results
to $\zeta_1$ and $\zeta_2$, the couplings of quintessence fields with
electromagnetism, and we have found that, in order to be consistent
with the data, these parameters must be at least one
order of magnitude smaller than the upper bound implied by the
Equivalence Principle.

On a more general ground, we could say that establishing whether there
is a variation of the fine structure constant and, in the affirmative
case, identifying its origin, remains a difficult task before a deeper
analysis of the systematic errors of the observations and studies of
the degeneracies with the various cosmological parameters.  This is
particularly evident in what concerns the compatibility of the
datasets of Murphy {\it et al.} and Chand {\it et al.}

\begin {acknowledgments}

\noindent
M.C. Bento and O. Bertolami acknowledge the partial support of
Funda\c c\~ao para a Ci\^encia e a Tecnologia (FCT) under the grant
POCTI/FIS/36285/2000.  N.M.C. Santos is supported by FCT grant
SFRH/BD/4797/2001.

\end{acknowledgments}

 
\end{document}